\documentclass[aps,notitlepage,nofootinbib,twocolumn]{revtex4-1}

\usepackage{amsmath,amssymb}
\usepackage{graphicx}

\newcommand{\eg}{e.g., }
\newcommand{\etaf}{\eta_{\rm f}}
\newcommand{\ie}{i.e., }
\newcommand{\kmax}{k_{\rm max}}
\newcommand{\tauc}{\tau_{\rm c}}
\newcommand{\xic}{\xi_{\rm c}}
\newcommand{\xis}{\xi_{\rm s}}

\begin{document}

\title{Criteria of amorphous solidification}

\author{Haim Diamant} 

\affiliation{Raymond and Beverly Sackler School of Chemistry, Tel
  Aviv University, Tel Aviv 6997801, Israel}

\begin{abstract}
A different perspective on the long-standing problem of amorphous
solidification is offered, based on an alternative definition of a
solid as a porous medium. General, model-free results are obtained
concerning the growing dynamic length accompanying solidification and
its relation to the growing relaxation time. Criteria are derived for
the dynamic length to diverge and for its divergence to entail the
arrest of particle motion.
\end{abstract}

\maketitle

\section{Introduction}

Despite the ubiquity of amorphous solids around us, the nature of the
transition from a liquid state to an amorphous solid (the glass
transition) remains one of the outstanding problems in
condensed-matter and chemical physics. After decades of research
\cite{DHbook,Binderbook,RitortSollich,MCReview,ParisiReview,ChandlerReview,Cates2010,DHReview,Weeks2012,Stillinger2013,Kaufman2013}
the situation is still confusing, especially in the theoretical
respect \cite{Stillinger2013,TarjusChapter}. It is unclear whether the
transition is caused by thermodynamic (entropic) effects or by kinetic
constraints, whether or not it is accompanied by a divergent length
scale, and what symmetry, if any, is broken when the material
solidifies. We shall not attempt to give here a balanced summary of
all existing theories; the reader is referred to a host of recently
published books and review articles on the subject
\cite{DHbook,Binderbook,RitortSollich,MCReview,ParisiReview,ChandlerReview,Cates2010,DHReview,Weeks2012,Stillinger2013,Kaufman2013}.
We note only that the various theories span a broad conceptual
range\,---\,from attempts to unravel {\it purely structural}
regularities in the disordered solid
\cite{KurchanLevine2009,KurchanLevine2011}, all the way to describing
the transition as a symmetry breaking along {\it purely temporal}
trajectories \cite{ChandlerReview}, and other approaches in-between
these two extremes.

In this state of affairs it seems constructive to pose a few modest
but essential questions and try to answer them as definitely as
possible {\it on a general, model-free level}. This is the goal of the
current contribution. The following analysis is based on the
assumption, which is in consensus, that amorphous solidification,
unlike crystallization, is a continuous transition, lacking any sharp
changes in the material's state variables. In particular, the
solidifying material is assumed to have a structural correlation
length, $\xis$, which does not appreciably change through the
transition. Based on this minimal premise, we shall obtain several
general criteria that are required for an amorphous solid to form.

\section{Growing dynamic length}
\label{sec_length}

Approaching amorphous solidification involves an enormous increase,
and possibly divergence, of the liquid's relaxation time,
$\tauc$. (Some have argued that an amorphous solid is nothing but a
tremendously slow liquid.\footnote{``The so-called amorphous solids
  are either not really amorphous or not really solid.''
  (E.\ Schr\"odinger) \cite{Schroedinger}.})  During the years there
has been an increasing recognition that the growing $\tauc$ is
accompanied by a growing dynamic length, $\xic$, corresponding to
cooperatively rearranging regions \cite{AdamGibbs}, or {\it dynamic
  heterogeneities} \cite{DHbook}. A few methods have been used to tap
into this length \cite{HarrowellChapter}, yielding the following
observations. The dynamic length increases together with the
relaxation time, albeit much more slowly\,---\,while $\tauc$ increases
by many orders of magnitude, $\xic$ increases by one order of
magnitude only. Sufficiently far from the transition, for
three-dimensional systems, the inter-dependence is found to be roughly
$\xic\sim\tauc^{1/z}$ with $z\simeq 2$
\cite{YamamotoOnuki1998,Lacevic2003}. Closer to the transition,
however, it becomes much weaker (possibly logarithmic)
\cite{Berthier2007,Flenner2010}, making it hard to determine whether
$\xic$ eventually diverges or not.

The first three questions that we pose, therefore, are as follows. (a)
Is a divergent dynamic length a necessary condition for amorphous
solidification? (b) How is it related to the growing time scale? (c)
What determines whether it diverges or not?

To answer these questions we first need a sharp distinction between
the two states that the transition separates, \ie a definition of a
solid (be it ordered or disordered). According to the old and useful
mechanical definition \cite{Alexander2}, a material is solid if it has
a finite, steady (zero-frequency), global (zero-wavevector), shear
modulus, $G(\omega=0,k=0)>0$. It follows that the material has a
divergent steady global shear viscosity,
$\eta(\omega,k)=G(\omega,k)/(i\omega) \xrightarrow{\omega,k\rightarrow
  0} \infty$.\footnote{It has recently been argued that the divergence
  of the viscosity, rather than the nonvanishing of the shear modulus,
  is the more accurate criterion of solidity \cite{Sausset2010}.} We
propose here an alternative definition:
\begin{quote}
{\it A material is solid if it acts as a stationary porous matrix on
  the flow of a fictitious viscous fluid filling the space between its
  constituent particles.}
\end{quote}  
We choose the fictitious continuum to be incompressible and
sufficiently fast-relaxing, such that it always reaches steady state
on the time scale of the actual material's relaxation. Accordingly, we
take the viscosity of the fluid to be much smaller than the material's
steady global viscosity, $\etaf\ll\eta(k=0)$. (The discussion focuses
hereafter on the zero-frequency limit; therefore, we omit it and keep
reference to the wavevector $k$ alone.) We include momentum exchange
between the two components via friction (\eg applying stick boundary
conditions) at the surfaces of the particles. To examine the state of
a material, consider the following thought experiment. Go to a volume
element of the pervading fluid, apply to it a steady point force $F$
(assumed sufficiently weak to cause a strictly linear response), and
measure the resulting steady fluid velocity $v$ at a very large
distance $r\gg\xis$. If the composite moves coherently there, the
fluid flowing together with the particulate matrix, the material is a
liquid; if the fluid moves relative to the matrix, the material is a
solid. This definition may seem awkward, but it will serve our purpose
well. In addition, (a) the alternative definition and the mechanical
one are equivalent, as shown below; and (b) the alternative definition
readily reveals a broken symmetry associated with amorphous
solidification\,---\,the Galilean invariance of the pervading fluid.

In more detail, up to the two leading terms in large $r/\xis$, the
fluid velocity at $r$ reads
\begin{eqnarray}
  v &=& v_1 + v_2 \nonumber\\
  v_1 &=& \frac{1}{4\pi\eta(0)r}F,\ \ \ 
  v_2 = \frac{\xis^2}{2\pi\etaf r^3}F,
\label{v}
\end{eqnarray}
where, for the sake of concreteness, we have taken $v$ as the fluid's
velocity component parallel to $F$. These expressions are of general
validity, holding for any viscous fluid embedded in a more rigid, but
still fluid, isotropic structure. They stem from the conservation laws
of momentum (of the entire composite) and mass (of the pervading
fluid) \cite{ijc07,jpsj09,prl14}. The first term, $v_1$, dominates at
very large distances. It describes the global response of the
composite as a whole to $F$. Its spatial $1/(\eta(0)r)$ decay follows
from conservation of the momentum emanating from the momentum source
$F$. The second term, $v_2$, is the leading correction to $v_1$. It
describes the flow of the fluid through the embedding structure at
smaller (but larger than $\xis$) distances. Its $\xis^2/(\etaf r^3)$
dependence arises from the displacement of fluid mass (effective mass
dipole), caused by $F$ within the ``pore'' size $\xis$. If the
structure becomes a stationary porous matrix for the embedded fluid,
only $v_2$ survives \cite{Brinkman,ijc07,jpsj09}, reflecting the loss
of translation invariance (equivalently, momentum conservation) of the
embedded fluid. Under the condition that $\xis$ remain finite, this
occurs if and only if $\eta(0)\rightarrow\infty$, as demanded by the
mechanical definition of a solid.\footnote{Another possibility to have
  $v_1$ lose ground to $v_2$ at large distances, is that $\xis$ grows
  indefinitely, as in a thermodynamic second-order phase
  transition. This does not correspond to solidification but to the
  critical slowing down of relaxation in such systems
  \cite{ChaikinLubensky}.}

The two terms in Eq.\ (\ref{v}) become equal at the distance
$\xic=\xis[2\eta(0)/\etaf]^{1/2}$. This distance separates two
different dynamic regions. In the ``porous region'', $\xis\ll
r\ll\xic$, the fluid flows relative to the matrix and imparts momentum
to it through friction. In the ``coherent region'', $r\gg\xic$, there
is bidirectional momentum exchange between the two components, and
they flow together as a momentum-conserving composite. The existence
of these two regions, together with the associated crossover distance,
has been demonstrated experimentally and theoretically for entangled
polymer networks in solution \cite{prl14}. (There is also a
small-scale, ``intra-pore'' region, $r\ll\xis$, where the fictitious
fluid flows without interacting with the matrix.)

\begin{figure}[h]
%\vspace{0.7cm} 
\centerline{\resizebox{8.5cm}{!}{\includegraphics{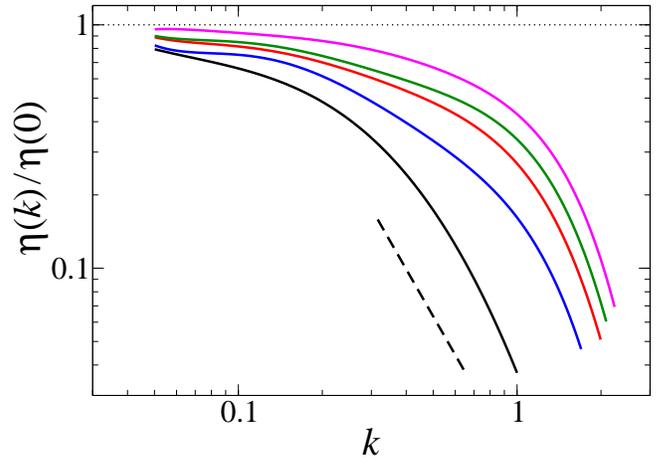}}}
\caption{Viscosity ratio as a function of wavevector, on a logarithmic
  scale, for a supercooled binary Lennard-Jones liquid. Different
  curves, from top to bottom, correspond to decreasing temperature
  toward the transition: $(T-T_{\rm c})/T_{\rm c} = 2.0$, $1.3$,
  $0.92$, $0.54$, and $0.15$. The dashed line has a slope of
  $-2$. Curves were adapted from Ref.\ \cite{KimKeyes2005}, Fig.\ 2.}
\label{figeta}
\end{figure}

To just determine whether the material is solid or liquid, according
to the symmetry breaking introduced above, we need not specify the
properties of the fictitious fluid. However, to get more details about
$\xic$, we must relate the hitherto unspecified $\etaf$ to the
material properties. The dynamics of the actual material has an upper
wavevector cutoff, $\kmax\sim\xis^{-1}$. Since the pervading fluid is
a continuum, the dynamics of the composite does not have this cutoff,
and in the ``intra-pore'' region $\eta(k>\kmax)=\etaf$. We thus set
$\etaf=\eta(k=\kmax)$ to get the simplest continuous extrapolation of
the material's viscosity to large wavevectors. (The requirement
$\etaf\ll\eta(0)$ is consequently translated to the requirement that
the small-scale viscosity of the material be much smaller than the
large-scale one, $\eta(\kmax)\ll\eta(0)$\,---\,a condition that is met
by glass-forming liquids; see, \eg Fig.\ \ref{figeta}.) This implies
the following relation for the dynamic length in terms of the actual
material properties:\footnote{Essentially, the fictitious fluid has
  been replaced now by the material's lower-viscosity, small-scale
  flow. Separating the solidifying material into two fluids is not a
  new idea \cite{Bendler1992,Schwartz}.}
\begin{equation}
  \xic = \xis \left[2\frac{\eta(0)}{\eta(\kmax)}\right]^{1/2} = 
  \xis \left[2\frac{\tauc}{\tau(\kmax)}\right]^{1/2},
\label{xic}
\end{equation}
where $\tauc$ is the global relaxation time, $\tau(\kmax)$ is the
relaxation time of structures of order $\xis$, and we have used the
proportionality between viscosity and relaxation time. We propose to
identify the length defined in Eq.\ (\ref{xic}) with the growing
length of solidification. Since we have assumed that $\xis$ does not
change discontinuously through the transition, $\xic$ increases
continuously with increasing
$\eta(0)/\eta(\kmax)=\tauc/\tau(\kmax)$. Physically, the length $\xic$
thus defined corresponds to the size of cooperative particle
structures, submerged in a background of noncooperative particles,
such that the two move relative to one another. In other words, $\xic$
is the characteristic size of a dynamic heterogeneity.

We are now faced with two possibilities, depending on the increase of
the large-scale relaxation time relative to the small-scale one. If
$\eta(0)$ increases more sharply than $\eta(\kmax)$ as the transition
is approached, then $\xic$ will diverge; if they become proportional
to one another (probably with a large proportionality factor), $\xic$
will remain finite. The former possibility corresponds to indefinitely
growing ``porous'' regions, with the small-scale dynamics remaining
liquid-like; the latter describes slowing down of the entire material
at all scales. As neither possibility can a priori be favored or
excluded, the divergence of $\tauc$ is not necessarily accompanied by
a divergent $\xic$.

Far from the transition, we expect from Eq.\ (\ref{xic}) to have
$\xic\sim\tauc^{1/z}$, $z=2$. As solidification is approached,
however, the relaxation at small length scales must become
increasingly slow too, making the ratio $\eta(0)/\eta(\kmax)$, and
with it $\xic$, increase much more moderately. These results are in
line with the observations mentioned above
\cite{YamamotoOnuki1998,Lacevic2003,Berthier2007}.

Let us examine the wavevector-dependent viscosity of the material for
$0<k<\kmax$. The function $\eta(k)/\eta(0)$ decreases from unity, its
value in the limit $k\rightarrow 0$, to
$(2\xis/\xic)^2\sim(\kmax\xic)^{-2}$, its value at $k=\kmax$ as given
by Eq.\ (\ref{xic}). There are two characteristic length scales, $\xis$
and $\xic$, of which $\xis$ provides the upper wavevector cutoff,
leaving us with $\xic$ alone. Thus, we can write in general,
$\eta(k)/\eta(0) = f(k\xic)$, where $f(x)$ is an unknown scaling
function. (The ability to collapse various wavevector-dependent
dynamic properties, once expressed as functions of $k\xic$, onto a
master curve has been demonstrated in various studies.) Assuming
$\xic\gg\xis$, we obtain two physically relevant ranges of $k$: (i)
$k\ll\xic^{-1}$; and (ii) $\xic^{-1}\ll k\leq\kmax$. These two ranges
correspond, respectively, to the large-distance,``coherent'' behavior
and the smaller-distance, ``porous'' one, discussed above.  To
interpolate between the two limits, $k=0$ and $k=\kmax$, without
introducing a third regime, we must set in range (ii) $f(k\xic)\sim
(k\xic)^{-2}$.  In summary, then,
\begin{equation}
  \frac{\eta(k)}{\eta(0)} \sim \left\{
  \begin{array}{ll}
    1, & k\ll\xic^{-1}\\
    (k\xic)^{-2},\ \ \  & \xic^{-1}\ll k<\kmax.
  \end{array} \right.
\label{eta_xic}
\end{equation}
In Fig. \ref{figeta} we reproduce simulation results for
$\eta(k)/\eta(0)$ in a supercooled binary Lennard-Jones liquid
\cite{KimKeyes2005}. As the transition is approached, the curves have
a broader range of $k$ where the viscosity ratio seems to scale as
$k^{-2}$, in accord with Eq.\ (\ref{eta_xic}).

\section{Motion arrest}
\label{sec_arrest}

From the preceding section we conclude that, if $\xic$ happens to
diverge, so does $\tauc$, and the material must solidify, whereas the
opposite direction is incorrect\,---\,if $\tauc$ diverges, a divergent
$\xic$ is not necessarily implied. 

There is yet another issue to consider. If a system exhibits
cooperative motion of arbitrarily many particles, must the associated
divergent length coincide with $\xic$ of the preceding section and
lead to solidification? That the answer to this question is negative
is evident from known counter-examples, of which we mention two: (a)
single-file diffusion\,---\,the random motion of particles along a
line without bypassing one another \cite{SFD}; and (b) the dynamics of
a long flexible polymer chain in solution \cite{DoiEdwards}. In both
examples the large-distance motion of a constituent particle requires
cooperative motion of an arbitrarily large number of other
particles. The particles in these systems are even
distinguishable\,---\,a broken symmetry previously suggested as a
criterion of solidity \cite{Alexander2,Alexander1,Schwartz}. Yet,
these systems are obviously not solid. Their cooperativity leads to
subdiffusion rather than freezing. In single-file diffusion the
mean-square displacement (MSD) of a particle scales with time as
$\langle(\Delta r)^2\rangle \sim t^{1/2}$; the MSD of a monomer in an
ideal polymer chain scales, respectively, as $t^{1/2}$ or $t^{2/3}$
for Rouse or Zimm dynamics (\ie excluding or including hydrodynamic
interactions). Thus, over a sufficiently long time, any particle in
these systems can get arbitrarily far away from its initial
position\,---\,a characteristic of a fluid, not a solid.

The natural fourth question, therefore, is what criterion an
indefinitely growing dynamic heterogeneity should satisfy to ensure
arrest of particle motion. To answer this last question we employ an
additional assumption\,---\,that the divergent dynamic length is
accompanied by {\it dynamic scaling}
\cite{ChaikinLubensky,Granek}\,---\,a property found also in the two
examples just mentioned.

Consider a system of particles, in which a single particle undergoes
subdiffusion with MSD $\langle (\Delta r)^2\rangle \sim t^{\alpha}$,
$0<\alpha< 1$. We assume that the subdiffusion is caused by
cooperative dynamics\,---\,i.e., at increasingly long times the motion
of a single particle is determined by the cooperative motion of an
increasingly large body of particles to which it belongs. Let us first
consider an {\it isolated} body of $N$ particles and spatial extent
$\xi_N\sim N^{1/d}$, where $d$ is the body's Hausdorff dimension. The
isolated body is embedded in a background of noncooperative
particles. Under these conditions, at sufficiently long time, $t_N$,
the random motion of a single particle in the body will cross over
from subdiffusion to the normal diffusion of the body's center of
mass. (Such a crossover is seen, for instance, in simulations of
supercooled liquids \cite{plateau}.) This will occur when
$t_N^{\alpha} \sim D_N t_N$, where $D_N$ is the center-of-mass
diffusion coefficient. Assuming that $D_N$ decreases with size as $D_N
\sim \xi_N^{-\beta}$, we get $t_N \sim \xi_N^{\beta/(1-\alpha)}$.

Now, instead of being isolated, let the body be part of an
indefinitely large cooperative system and interact with its
surroundings over a relaxation time $\tau_N \sim \xi_N^{z}$. We demand
that there be no crossover time for this arbitrary size $N$, such that
the single-particle subdiffusion should persist at all times. We
therefore equate $t_N\sim\tau_N$, which leads to the relation,
\begin{equation}
  \alpha = 1 - \beta/z.
\label{alpha}
\end{equation}

Before turning to the case of solidification, let us first check the
validity of this scaling scheme for the cooperative but fluid systems
mentioned above. In single-file diffusion, $d=1$, $D_N \sim
N^{-1}\sim\xi_N^{-1}$, and $\tau_N\sim\xi_N^2$. Thus, $\beta=1$ and
$z=2$, leading, according to Eq.\ (\ref{alpha}), to $\alpha=1/2$. For
an ideal polymer with Rouse dynamics, we have $d=2$, $D_N \sim
N^{-1}\sim \xi_N^{-2}$, and $\tau_N\sim\xi_N^2/D_N \sim
\xi_N^4$\,---\,\ie $\beta=2$ and $z=4$. This leads again, but for
different reasons, to $\alpha=1/2$. For an ideal polymer with Zimm
dynamics, $D_N\sim\xi_N^{-1}$ and $\tau_N\sim\xi_N^2/D_N\sim
\xi_N^{3}$. The resulting $\beta=1$ and $z=3$ yield $\alpha=2/3$. Thus,
the known subdiffusion exponents are correctly reproduced. (One can
verify this scheme for other examples as well, \eg self-avoiding
polymers.)

In the case of solidification, which necessitates the formation of a
rigid reference frame of particle positions \cite{Alexander2}, we
require (on the ensemble-average level) that the long-distance motion
of particles be arrested, \ie $\alpha\rightarrow 0$
\cite{plateau}.\footnote{Selecting $\alpha\rightarrow 0$ includes also
  logarithmic creeping.} From Eq.\ (\ref{alpha}) we then get the
criterion of solidification as $\beta=z$, which is equivalent to
$D_N\tau_N \sim N^0\sim l^2$, $l$ being the size of a particle. The
physical meaning of this result is that, as the material solidifies,
the center-of-mass displacement of an arbitrarily sized heterogeneity
before it relaxes does not extend to more than a few particle sizes.

Utilizing once again the proportionality between relaxation time and
viscosity, we have
\begin{equation}
  \frac{\eta(k)}{\eta(0)} = \frac{\tau_N(k\sim\xi_N^{-1})} 
  {\tau_N(k\sim\xic^{-1}\rightarrow 0)} \sim (\xic k)^{-z}.
\label{eta_scaling}
\end{equation}
Comparison of Eqs.\ (\ref{eta_xic}) and (\ref{eta_scaling}) leads to
the more precise criterion,
\begin{equation}
  \beta = z = 2,
\label{z}
\end{equation}
the fulfillment of which ensures that the divergent cooperativity
length coincide with the solidification length of the preceding
section and lead to arrest.

Furthermore, we claim that the center-of-mass diffusion coefficient of
a heterogeneity is inversely proportional to the number of particles
that it contains, $D_N\sim N^{-1}$. Such Rouse-like dynamics takes
place when the friction with the environment, experienced by each
particle, is independent of the other particles in the object. This
happens, in turn, when the propagation of stresses in the environment
across the object is not fast enough relative to the object's
motion. In ordinary circumstances, where a body moves through a
fast-relaxing fluid, the opposite holds. The time scale of stress
propagation is $\xi_N^2/\nu$, $\nu$ being the fluid's kinematic
viscosity, and the one related to the object's motion is
$\xi_N^2/D_N$. Their ratio is $D_N/\nu$, which is very small unless
the object is of atomic dimensions. Consequently, in this case
$D_N\sim\xi_N^{-1}$, as in the Stokes-Einstein formula or the Zimm
dynamics of a polymer. In our case the stress-propagation time is the
same, $\xi_N^2/\nu$, but the object's motion is much more
restricted. According to the foregoing discussion, the latter time
scale is only $\tau_N\sim l^2/D_N$. The ratio, therefore, is
$(\xi_N/l)^2(D_N/\nu)\sim (l/\xi_N)^{\beta-2}$. Since $\beta=2$, there
is no separation of time scales, and $D_N\sim N^{-1}\sim
\xi_N^{-d}$. Equation (\ref{z}) then yields $d=2$. A fractal dimension
of 2 has quite consistently been observed for dynamic heterogeneities
in several glassy systems
\cite{Weeks2000,Castillo2003,Giovambattista2005,Zhang2011,Starr2013}.

\section{Conclusion}

The conditional arguments developed here are not to be mistaken for a
theory of the glass transition. For example, they do not provide
information as to how the dynamic heterogeneities form, the specific
properties that lead to their fractal dimension, the dependence of the
transition properties on control parameters such as temperature or
density, the actual value of the ratio $\eta(0)/\eta(\kmax)$
determining the divergence of $\xic$, and whether all these and other
features are universal or differ from one system to the next.

Nevertheless, these arguments have yielded several concrete
predictions, which we now wish to summarize. (a) The growing length
scale in amorphous solidification depends on the (square-root of) the
ratio between the material's largest- and smallest-scale relaxation
times. Consequently, it may or may not diverge, depending on the
asymptotic behavior of that ratio as the transition is
approached. This finding implies also that different glassy systems
may behave qualitatively differently. (b) For $k>\xic^{-1}$ the
wavevector-dependent viscosity (and relaxation time) scales as
$\eta(k)/\eta(0)\sim(k\xic)^{-z}$ with $z=2$. (c) In case of a
divergent $\xic$, the dynamic exponents, characterizing the relaxation
time and center-of-mass diffusion coefficient of dynamic
heterogeneities, are equal, $z=\beta=2$, and the heterogeneities'
Hausdorff dimension is 2.

Some of these predictions are in line with available experimental and
numerical data, while others call for further study\,---\,in
particular, those pertaining to the divergence of the dynamic
length. On the one hand, the successful dynamic scaling might indicate
a divergent $\xic$. On the other hand, if the transition is found to
be a genuine dynamic criticality, a more accurate analysis (\eg
concerning the critical exponents) will be called for. Most of all, we
hope that the general results obtained here are found useful when a
widely accepted molecular theory of amorphous solidification emerges
and its validity is checked.

\begin{acknowledgments}
I am grateful to A.\ Sonn-Segev and Y.\ Roichman, whose experiments
inspired part of this analysis. This research was supported by the
Israel Science Foundation under Grant No.\ 8/10.
\end{acknowledgments}


\begin{thebibliography}{99}

\bibitem{DHbook} 
Berthier L, Biroli G, Bouchaud J-P, Cipelleti L, 
van~Saarloos W (2011)
{\it Dynamical Heterogeneities in Glasses, Colloids, and Granular Media}
(Oxford University Press).

\bibitem{Binderbook}
Binder K, Kob W (2011)
{\em Glassy Materials and Disordered Solids: An Introduction to Their 
Statistical Mechanics} (World Scientific).

\bibitem{RitortSollich}
Ritort F, Sollich P (2003)
Glassy dynamics of kinetically constrained models.
{\it Adv Phys} 52:219-342.

\bibitem{MCReview}
Das SP (2004)
Mode-coupling theory and the glass transition in supercooled liquids.
{\it Rev Mod Phys} 76:785-851.

\bibitem{ParisiReview}
Parisi G, Zamponi F (2010) 
Mean-field theory of hard sphere glasses and jamming.
{\it Rev Mod Phys} 82:789-845.

\bibitem{ChandlerReview}
Chandler D, Garrahan JP (2010)
Dynamics on the way to forming glass: Bubbles
in space-time.
{\it Annu Rev Phys Chem} 61:191-217.

\bibitem{Cates2010}
Cates ME (2010)
Mode-coupling theory for the rheology of colloidal glasses:
Recent progress.
{\it Progr Theo Phys Suppl} 184:222-231.

\bibitem{DHReview}
Berthier L, Biroli G (2011)
Theoretical perspective on the glass transition and amorphous materials.
{\it Rev Mod Phys} 83:587-645.

\bibitem{Weeks2012}
G.\ L.\ Hunter GL, Weeks ER (2012)
The physics of the colloidal glass transition.
{\it Rep Progr Phys} 75:066501. 

\bibitem{Stillinger2013}
Stillinger FH, Debenedetti PG (2013)
Glass transition thermodynamics and kinetics.
{\it Annu Rev Condensed Matter Phys} 4: 263-285.
% Review of controversy in theories

\bibitem{Kaufman2013}
Kaufman LJ (2013)
Heterogeneity in single-molecule observables in the study of
 supercooled liquids.
{\it Annu Rev Phys Chem} 64:177-200.

%\bibitem{Szamel2013}
%Szamel G (2013)
%Mode-coupling theory and beyond: A diagrammatic approach.
%{\it Progr.\ Theo.\ Exp.\ Phys.} 012J01.

\bibitem{TarjusChapter}
Tarjus D (2011) 
An overview of the theories of the glass transition.
in \cite{DHbook}, pp 39-67.

\bibitem{KurchanLevine2009}
Kurchan J, Levine D (2009)
Correlation length for amorphous systems.
arXiv:0904.4850.

\bibitem{KurchanLevine2011}
Kurchan J, Levine D (2011)
Order in glassy systems.
{\it J Phys A} 44:035001.

\bibitem{Schroedinger} 
Schr\"odinger E (1967), {\it What is Life} (Cambridge University Press), 
p 59.

\bibitem{AdamGibbs}
Adam G, Gibbs JH (1965)
On the temperature dependence of cooperative relaxation properties in
glass-forming liquids.
{\it J Chem Phys} 43:139-146.

\bibitem{HarrowellChapter}
Harrowell P (2011)
The length scales of dynamic heterogeneity: Results from molecular 
dynamics simulations. in \cite{DHbook}, pp 229-263. 

\bibitem{YamamotoOnuki1998}
Yamamoto R, Onuki A (1998)
Dynamics of highly supercooled liquids: Heterogeneity, rheology, 
and diffusion.
{\it Phys Rev E} 58:3515-3529.

\bibitem{Lacevic2003}
La\v{c}evi\'{c} N, 
Starr FW, Schr{\o}der TB, Glotzer SC (2003)
Spatially heterogeneous dynamics investigated via a time-dependent 
four-point density correlation function.
{\it J Chem Phys} 119:7372-7387.

\bibitem{Berthier2007}
Berthier L, Biroli G, Bouchaud J-P, Kob W, Miyazaki K, Reichman D (2007)
Spontaneous and induced dynamic correlations in glass formers II
Model calculations and comparison to numerical simulations.
{\it J Chem Phys} 126:184504.

\bibitem{Flenner2010}
Flenner E, Szamel G (2010)
Dynamic heterogeneity in a glass forming fluid: Susceptibility,
structure factor, and correlation length.
{\it Phys Rev Lett} 105:217801.

\bibitem{Alexander2}
Alexander S (1998)
Amorphous solids: Their structure, lattice dynamics and elasticity.
{\it Phys Rep} 296:65-236.

\bibitem{Alexander1}
Alexander S (1998)
What is a solid?.
{\it Physica A} 249:266-275.

\bibitem{Sausset2010}
Sausset F, Biroli G, Kurchan J (2010)
Do solids flow?.
{\it J Stat Phys} 140:718-727.

\bibitem{ijc07}
Diamant H (2007)
Long-range hydrodynamic response of particulate liquids
and liquid-laden solids.
{\it Isr J Chem} 47:225-231.

\bibitem{jpsj09}
Diamant H (2009)
Hydrodynamic interaction in confined geometries.
{\it J Phys Soc Jpn} 78:041002.

\bibitem{prl14}
Sonn-Segev A, Bernheim-Groswasser A, Diamant H, Roichman Y (2014)
Viscoelastic response of a complex fluid at intermediate distances.
{\it Phys Rev Lett} 112:088301.

\bibitem{Brinkman}
Brinkman HC (1947)
A calculation of the viscous force exerted by a flowing fluid
on a dense swarm of particles.
{\it Appl Sci Res} A1:27-34.

\bibitem{ChaikinLubensky}
Chaikin PM, Lubensky TC (1995)
{\it Principles of Condensed Matter Physics}
(Cambridge University Press).

\bibitem{Bendler1992}
Bendler JT, Shlesinger MF (1992)
Defect diffusion and a two-fluid model for structural relaxation near 
the glass-liquid transition.
{\it J Phys Chem} 96:3970-3973.

\bibitem{Schwartz}
Schwartz M (2004)
Bose-Einstein condensation and the glassy state.
{\it Phys Rev Lett} 93:205701.

\bibitem{KimKeyes2005}
Kim J, Keyes T (2005)
On the breakdown of the Stokes-Einstein law in supercooled liquids.
{\it J Phys Chem B} 109:21445-21448.

\bibitem{SFD}
Alexander S, Pincus P (1978)
Diffusion of labeled particles on one-dimensional chains.
{\it Phys Rev B} 18:2011-2012.

\bibitem{DoiEdwards}
Doi M, Edwards SF (1986)
{\it The Theory of Polymer Dynamics}
(Oxford University Press).

\bibitem{Granek}
Reuveni S, Klafter J, Granek R (2012)
Dynamic structure factor of vibrating fractals.
{\it Phys Rev Lett} 108, 068101.

\bibitem{plateau}
Szamel G, Flenner E (2006)
Time scale for the onset of Fickian diffusion in supercooled liquids.
{\it Phys Rev E} 73:011504.

\bibitem{Weeks2000}
Weeks ER, Crocker JC, Levitt AC, Schoefield A, Weitz DA (2000)
Three-dimensional direct imaging of structural relaxation 
near the colloidal glass transition.
{\it Science} 287:627-631.

\bibitem{Castillo2003}
Castillo HE, Chamon C, Cugliandolo LF, Iguain JL, Kennett MP (2003) 
Spatially heterogeneous ages in glassy systems.
{\it Phys Rev B} 68:134442.

\bibitem{Giovambattista2005} 
Giovambattista N, Buldyrev SV, Starr FW, Stanley HE (2005)
Clusters of mobile molecules in supercooled water.
{\it Phys Rev E} 72: 011202 (2005).

\bibitem{Zhang2011}
Zhang Z, Yunker PJ, Habdas P, Yodh AG (2011)
Cooperative rearrangement regions and dynamical heterogeneities 
in colloidal glasses with attractive versus repulsive interactions.
{\it Phys Rev Lett} 107:208303.

\bibitem{Starr2013}
Starr FW, Douglas JF, Sastry S (2013)
The relationship of dynamical heterogeneity to the Adam-Gibbs
and random first-order transition theories of glass formation.
{\it J Chem Phys} 138:12A541.

\end{thebibliography}
\end{document}